\documentclass[doublecol]{epl2} 

\usepackage{amssymb,amsfonts,bm}
\usepackage{graphics,graphicx}

\title{Ground state structure of a bilayer Wigner crystal with
repulsive dielectric images}
\shorttitle{Ground state of bilayer Wigner crystals with repulsive images} 

\author{Ladislav \v{S}amaj\inst{1,2} \and Emmanuel Trizac\inst{1}}
\shortauthor{Ladislav \v{S}amaj and Emmanuel Trizac}

\institute{                    
\inst{1} Universit\'e Paris-Sud, Laboratoire de Physique Th\'eorique et 
Mod\`eles Statistiques, UMR CNRS 8626, \\ 91405 Orsay, France, EU \\
\inst{2} Institute of Physics, Slovak Academy of Sciences, D\'ubravsk\'a
cesta 9, 845 11 Bratislava, Slovakia, EU}


\abstract{We study the ground-state structures of identical classical point 
charges with Coulomb interactions, confined between two symmetric parallel 
charged walls.
For the well understood homogeneous dielectric case with no electrostatic 
images, the charges evenly condense on the opposite walls, thereby forming 
a bilayer Wigner crystal; five structures compete upon changing 
the inter-wall separation. 
Here, we consider a dielectric jump between the walls and a solvent 
in which charges are immersed, implying repulsive images.
Using recently developed series representations of lattice sums 
for Coulomb law, we derive the complete phase diagram.
In contrast to the homogeneous dielectric case, the particles remain in
a hexagonal Wigner monolayer up to a certain distance between the walls.
Beyond this distance, a bifurcation occurs to a sequence of Wigner bilayers,
each layer having a nonzero spacing from the nearest wall. 
Another new phenomenon is that the ground-state energy as a function of 
the wall separation exhibits a global minimum.}

\begin{document}

\maketitle

\section{Introduction}
Bilayer Wigner crystals of electrons appear in various experimental settings: 
GaAs quantum wells \cite{Manoharan96,Tutuc04} or other semiconductors 
\cite{Eisenstein04}, quantum dots \cite{Imamura96}, dusty plasmas 
\cite{Teng03}, colloids \cite{Neser97}, etc.
Theoretically, it is relevant and non-trivial 
to understand the ground-state features of 
Coulombic bilayers, starting with the classical limit. 
This subject in its homogeneous dielectric version has received significant 
attention in last years, as such 
\cite{Falko94,Esfarjani95,Goldoni96,Messina03,Oguz09,Lobaskin07} 
or within more general finite temperature analysis 
\cite{Schweigert99a,Schweigert99b,Weis01,Mazars11}. 

\begin{figure}[htbp]
\begin{center}
\onefigure[width=0.39\textwidth]{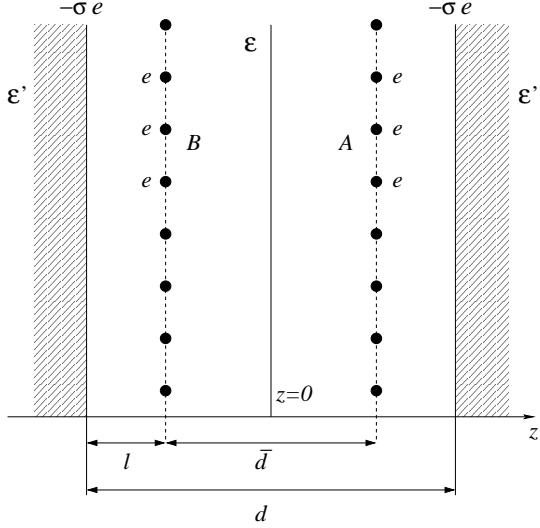}
\caption{The geometry of two dielectric walls at distance $d$, 
carrying the surface charge density $-\sigma e$.
The particles of charge $e$ form a Wigner bilayer which consists of 
two sub-lattices $A$ and $B$ at distance $\bar{d}$; 
a monolayer corresponds to $\bar{d}=0$.
The distance of each sub-lattice to the nearest wall is $l=(d-\bar{d})/2$.}
\label{fig:geometry}
\end{center}
\end{figure}

We consider the ground state of an ensemble of (say elementary) classical 
point charges $e$, interacting through the $1/r$ usual Coulomb pair potential.
The charges are confined between two symmetric parallel walls at distance $d$,
carrying a uniform surface charge of density $-\sigma e$ 
(see fig. \ref{fig:geometry}).
The system as a whole is electroneutral.
The static dielectric constant of the walls $\varepsilon'$ can be, in general, 
different from the constant $\varepsilon$ of the solvent in 
which the charges are immersed.
The dielectric jump between the two moieties is defined as 
$\Delta=(\varepsilon-\varepsilon')/(\varepsilon+\varepsilon')$, so that
$\Delta\in [-1,1]$.
The interval $\Delta\in [-1,0)$ corresponds to attractive and
$\Delta\in (0,1]$ to repulsive electrostatic image charges.
In realistic systems, in particular those of biological interest, 
the wall mimics the interior
of a polarizable colloid with $\varepsilon'\le 10$ while $\varepsilon\simeq 80$
for water solvent, so $\Delta$ is close to 1.

\begin{figure}[htbp]
\begin{center}
\onefigure[width=0.42\textwidth]{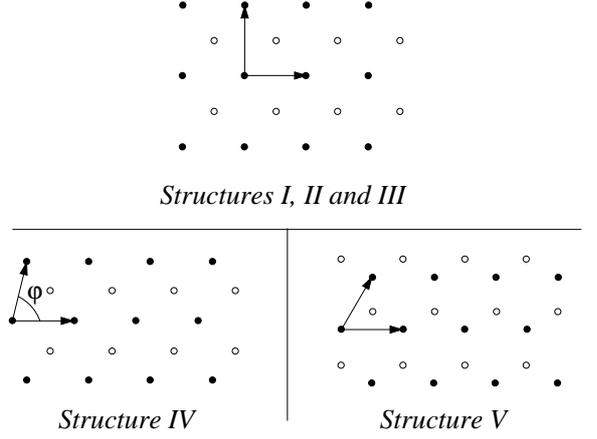}
\caption{Schematic representation of the different ground states 
encountered when the dimensionless distance $\eta$ increases,
in the homogeneous dielectric case $\Delta=0$.
The open and filled symbols show the locations of ions on 
the opposite surfaces.
The arrows are for the lattice vectors $\bm{a}_1$ and $\bm{a}_2$, 
from which we define the aspect ratio 
$\delta=\vert\bm{a}_2\vert/\vert\bm{a}_1\vert$: $\delta=\sqrt 3$ 
for structure I, $\delta=1$ for structure III, while structure II
is intermediate, having $1<\delta<\sqrt 3$. 
For structure IV, the order parameter is the angle $\varphi$ 
between $\bm{a}_1$ and $\bm{a}_2$. Structures IV and V have $\delta=1$.} 
\label{fig:struct}
\end{center}
\end{figure}

In the well understood homogeneous dielectric case $\Delta=0$, 
Earnshaw theorem \cite{Earnshaw} tells us that in the ground state,
the charges evenly condense on the opposite walls.
The structure of the formed bilayer depends on a single dimensionless 
parameter $\eta = d \sqrt{\sigma}$.
For $\eta=0$, a genuine hexagonal Wigner crystal (structure I) is stable
(provides the true minimum of the energy); this structure becomes unstable 
for an arbitrary small $\eta>0$ \cite{Oguz09,Samaj12a,Samaj12b}. 
In the opposite limit $\eta \to \infty$, the two layers decouple and 
a hexagonal Wigner crystal is formed on each plate (structure V);
to minimize the inter-layer repulsion, these two crystals adopt 
a staggered configuration.
For intermediate values of $\eta$, three other structures are met, see
fig. \ref{fig:struct}: a staggered rectangular lattice (structure II),
a staggered square lattice (structure III) and a staggered rhombic lattice
(structure IV) \cite{Goldoni96}.
The transitions between structures (II,III) and (III,IV) are 
of second order with mean-field critical indices \cite{Samaj12a,Samaj12b}.
The transition between structures (IV,V) is discontinuous, characterized
by a skip of the angle $\varphi$ to $\pi/3$ and by a mutual shift of
the lattices on the opposite surfaces.

In this letter, we consider the dielectric jump $\Delta\in [0,1]$,
implying repulsive electrostatic image charges.
The presence of repulsive image and self-image forces has intuitively
two important effects on the ground state:
\begin{itemize}
\item
the particles no longer collapse on the surface of the walls, but
create a symmetric bilayer structure with a nonzero spacing from
the walls, i.e. $\bar{d}\ne d$ in fig. \ref{fig:geometry};
\item 
the intervals of $\eta$, inside which the structures I-V are stable,
are changed.
In particular, a monolayer with structure I is expected to be stable up to 
a critical value $\eta_{\,\rm I\to II}>0$ beyond which it bifurcates 
into a bilayer with structure II. 
\end{itemize}

The nonzero spacing from the walls is confirmed in the limit 
$\eta\to\infty$ where the two layers decouple (up to a staggered 
shift) and each of them forms a neutral entity with 
the corresponding charged wall.
For the resulting one-wall geometry, it was shown in ref. \cite{Samaj12c} that, 
as a result of balance between attractive surface charge and repulsive 
image charge forces, the stable hexagonal Wigner crystal of 
particles is formed at some distance $l$ from the wall.
This distance is given by $l=at/2$, where $a$ is the lattice spacing
of the two-dimensional Wigner crystal determined by the electroneutrality
condition as $1 = (\sqrt{3}/2)a^2\sigma$ and the dimensionless parameter $t$
is the solution of the transcendental equation
\begin{equation} \label{t}
\sum_{j,k=-\infty}^{\infty} \frac{1}{(t^2+j^2+jk+k^2)^{3/2}}
= \frac{4\pi}{\sqrt{3}} \frac{1+\Delta}{\Delta} \frac{1}{t} .
\end{equation}  
The lattice sum on the lhs can be expressed as
\begin{equation} \label{tt}
\frac{2}{\pi} \int_0^{\infty} {\rm d}u \sqrt{u} {\rm e}^{-u t^2}
\left[ \theta_3({\rm e}^{-u}) \theta_3({\rm e}^{-3u})
+ \theta_2({\rm e}^{-u}) \theta_2({\rm e}^{-3u}) \right] ,
\end{equation}
where $\theta_2(q)=\sum_{n=-\infty}^{\infty} q^{\left( n-\frac{1}{2}\right)^2}$
and $\theta_3(q)=\sum_{n=-\infty}^{\infty} q^{n^2}$ are the Jacobi theta
functions with zero argument.

\section{Method}
Our goal is to calculate the Coulomb interaction energy of 
the electroneutral system pictured in fig. \ref{fig:geometry}.
Each point will be represented by Cartesian coordinates as 
${\bf r}=({\bf R},z)$, where ${\bf R}=(x,y)$ and the $z$ axis 
is perpendicular to the walls. 
The system is inhomogeneous along the $z$-axis, symmetric with respect to 
the $z=0$ plane.
The wall surfaces, charged uniformly by $-\sigma e$, are localized 
at $z=\pm d/2$.
The two Wigner sublayers, each formed of $N/2$ point charges $e$, 
are localized at $z=\bar{d}/2$ 
(sub-lattice $A$ with sites $\{ {\bf R}_j^A \}$) and at
$z=-\bar{d}/2$ (sub-lattice $B$ with sites $\{ {\bf R}_j^B \}$).
The distance of the Wigner sub-lattice $A$ (resp. $B$) to the nearest wall 
surface at $z=+d/2$ ({resp.} $z=-d/2$) is $l= (d-\bar{d})/2$.
We have $l>0$ for the considered case with repulsive images 
$\Delta\in (0,1]$ and $l=0$ otherwise (for attractive images 
$\Delta\in [-1,0)$, a small hard core around the particles is necessary 
to prevent thermodynamic collapse onto their images).

The Coulomb energy consists of three contributions:

\noindent {\bf (i) Particle-particle interactions}:
The direct Coulomb potential between two points ${\bf r}=({\bf R},z)$ 
and ${\bf r}'=({\bf R}',z')$, situated in between the walls, is given by
\begin{equation}
u_0({\bf r},{\bf r}') = \frac{1}{\varepsilon} 
\frac{1}{\sqrt{\vert {\bf R}-{\bf R}'\vert^2+(z-z')^2}} .
\end{equation}
Due to the dielectric inhomogeneity, a unit charge at ${\bf r}=({\bf R},z)$ 
has an infinite number of electrostatic images with the same vector ${\bf R}$ 
\cite{Jackson}. 
The first sequence of images, generated by considering first the reflection 
with respect to the wall surface at $z=+d/2$, is localized at 
$(-1)^{n+1}(nd-z)$ with associated charges $\Delta^n$ $(n=1,2\ldots)$. 
The second sequence, generated by considering first the reflection with respect
to the wall surface at $z=-d/2$, is localized at $(-1)^n(nd+z)$ with
associated charges $\Delta^n$ $(n=1,2\ldots)$. 
The image Coulomb potential thus reads
\begin{eqnarray}
u_{\rm im}({\bf r},{\bf r}') & = & \frac{1}{\varepsilon} \sum_{n=1}^{\infty} 
\Bigg\{ 
\frac{\Delta^{2n}}{\sqrt{\vert {\bf R}-{\bf R}'\vert^2+(2nd+z-z')^2}} 
\nonumber \\ & &
+ \frac{\Delta^{2n}}{\sqrt{\vert {\bf R}-{\bf R}'\vert^2+(2nd-z+z')^2}}
\nonumber \\ & &
+ \frac{\Delta^{2n-1}}{\sqrt{\vert {\bf R}-{\bf R}'\vert^2+[(2n-1)d+z+z']^2}}  
\nonumber \\ & &
+ \frac{\Delta^{2n-1}}{\sqrt{\vert {\bf R}-{\bf R}'\vert^2+[(2n-1)d-z-z']^2}}  
\Bigg\} . \nonumber \\ & & \label{images}
\end{eqnarray}
The notation 
$u_{\rm im}({\bf r},{\bf r}')\equiv u_{\rm im}(\vert {\bf R}-{\bf R}'\vert;z,z')$
will be often used in what follows. 
The total energy of particles and their images is given by
\begin{equation}
E_{pp} = \frac{e^2}{2} \left[ \sum_{j\ne k} u_0({\bf r}_j,{\bf r}_k)
+ \sum_{j,k} u_{\rm im}({\bf r}_j,{\bf r}_k) \right] .
\end{equation}
The term $j=k$ in the second sum corresponds to the interaction
of particle $j$ with its own image. 
Let us {choose} as a reference particle the one at site $1$ of 
lattice $A$ localized on the plane $z=\bar{d}/2$.
The energy per particle is expressible as
\begin{eqnarray}
\frac{E_{pp}}{N} & = & \frac{e^2}{2\varepsilon} \left[
\sum_{j\ne 1} \frac{1}{R_{1j}^{AA}} 
+ \sum_j \frac{1}{\sqrt{(R_{1j}^{AB})^2+\bar{d}^2}} \right] \nonumber \\
& & + \frac{e^2}{2} \left[ 
u_{\rm im}\left( 0;\frac{\bar{d}}{2},\frac{\bar{d}}{2}\right) + \sum_{j\ne 1} 
u_{\rm im}\left( R_{1j}^{AA};\frac{\bar{d}}{2},\frac{\bar{d}}{2}\right) 
\right. \nonumber \\ & & \left. + \sum_j 
u_{\rm im}\left( R_{1j}^{AB};\frac{\bar{d}}{2},-\frac{\bar{d}}{2}\right)
\right] . \label{latticesums}
\end{eqnarray}
Here, $R_{1j}^{AA}\equiv \vert {\bf R}_1^A-{\bf R}_j^A\vert$ and similarly
$R_{1j}^{AB}\equiv \vert {\bf R}_1^A-{\bf R}_j^B\vert$.
Owing to the symmetry of the problem, the same formula holds
if the reference particle sits on lattice $B$.

The lattice sums in the above expression have to be ``regularized'' 
by background terms in the following sense.
Let us consider two plates at distance $z$, each of a large surface $S$.
They carry the fixed surface charge density $-\sigma e$.
At each plate, there are point charges $e$ of surface density $\sigma$.
The total number of particles $N=2\sigma S$, so that the system as a whole 
is electroneutral.
The dielectric constant is homogeneous, equal to $\varepsilon$.
The energy per unit surface due to the background-background interaction of 
the two $(-\sigma e)$ plates at distance $z$ is given by 
$E_{bb}/S = -2\pi (\sigma e)^2 z/\varepsilon$. 
Equivalently, $E_{bb}/N = -\pi \sigma e^2 z/\varepsilon$. 
The interaction energy of charge $e$ with $(-\sigma e)$ plate at 
distance $z$ is $2\pi\sigma e^2 z/\varepsilon$.
The total ``background energy'' per particle thus reads
$\epsilon(z) = \pi\sigma e^2 z/\varepsilon$.
Such term has to be added to and subtracted from each of the lattice sums 
in (\ref{latticesums}).
The first term is unchanged, the second term becomes 
$$
\frac{e^2}{2\varepsilon} 
\left[ \sum_j \frac{1}{\sqrt{(R_{1j}^{AB})^2+\bar{d}^2}}
+ 2\pi\sigma \bar{d} \right] - \frac{\pi\sigma e^2}{\varepsilon} \bar{d} , 
$$
etc.
The regularized sums (like the one in the square bracket) vanish in 
the large-distance limit. Interestingly, they are amenable to
a special representation by quickly converging 
series expansions, with the aid of the recent method 
developed in \cite{Samaj12a,Samaj12b}. 
Note that the regularization of image interactions $u_{\rm im}$ requires 
addition and subtraction of an infinite number of background energies.
After performing the regularization of all lattice sums 
in (\ref{latticesums}), we end up with 
\begin{equation}
\frac{E_{pp}}{N} = \frac{E^*_{pp}}{N} - \frac{\pi\sigma e^2}{\varepsilon} \bar{d}
- \frac{4\pi\sigma e^2}{\varepsilon}  \frac{\Delta}{(1-\Delta)^2} d ,
\end{equation}
where the star means ``regularized''.

\noindent {\bf (ii) Particle-surface charge interaction}:
We now study the direct and image interactions of charged particles with 
the fixed surface charge densities on the walls.
The surface charge $-\sigma e$ at $z=-d/2$ has one image $\Delta(-\sigma e)$ 
at the same $z=-d/2$ and a series of images $(1+\Delta)\Delta^{n-1}(-\sigma e)$ 
at $z=(-1)^n (2n-1) d/2$ $(n=2,3,\ldots)$.
The surface charge $-\sigma e$ at $z=+d/2$ has one image $\Delta(-\sigma e)$ 
at the same $z=+d/2$ and a series of images $(1+\Delta)\Delta^{n-1}(-\sigma e)$ 
at $z=(-1)^{n-1} (2n-1) d/2$ $(n=2,3,\ldots)$.
After simple algebra, the interaction energy of one particle of charge $e$ 
with both surface charges $(-\sigma e)$ and their images takes the form
\begin{equation} \label{ps}
\frac{E_{ps}}{N} = \frac{2\pi\sigma e^2}{\varepsilon}  
\frac{(1+\Delta)^2}{(1-\Delta)^2} d .
\end{equation}
The same result is obtained, as it should be, when one computes 
the interaction energy of both surface charges $(-\sigma e)$ 
with one charge $e$ and all its images.

\noindent {\bf (iii) Surface charge-surface charge interaction}:
The energy per particle due to the direct and image interaction 
between surface charges turns out to be
\begin{equation}
\frac{E_{ss}}{N} = - \frac{\pi\sigma e^2}{\varepsilon}  
\frac{(1+\Delta)^2}{(1-\Delta)^2} d .
\end{equation}
Note that this value equals to $-1/2$ of $E_{ps}/N$ defined by (\ref{ps}), 
in close analogy with the homogeneous $\Delta=0$ case.

For the total energy $E=E_{pp}+E_{ps}+E_{ss}$, we get
\begin{equation} \label{energy}
\frac{E}{N} = \frac{E^*_{pp}}{N} + 
\frac{\pi\sigma e^2}{\varepsilon} (d - \bar{d}) .
\end{equation}
The infinite lattice sums in $E^*_{pp}/N$, which is the regularized version 
of (\ref{latticesums}), can be expressed as quickly converging series of 
the special functions \cite{Samaj12a,Samaj12b}
\begin{equation} \label{znu}
z_{\nu}(x,y) = \int_0^{1/\pi} \frac{{\rm d}t}{t^{\nu}} {\rm e}^{-xt}
{\rm e}^{-y/t} , \qquad y>0 ,
\end{equation} 
which represent a generalization to two-layer problems of the so-called 
Misra functions \cite{Born40} used in single-layer lattice summations 
\cite{Bowick06}.
The cutoff of the series at the 5th term corresponds to the precision
of more than 17 decimal digits; 
we keep this cutoff in the present work.
As concerns the number of counted images in the image Coulomb potential 
(\ref{images}), we cut the series at $n=5$ as well; there exist various
checks indicating that this cutoff preserves excellent accuracy
of our results.

We do not assume that the inclusion of repulsive images will lead 
to the appearance of a new structure, different from those presented 
in fig. \ref{fig:struct}. 
The energy per particle of a structure has to be minimized with 
respect to all structure parameters.
In comparison with the $\Delta=0$ case, 
the bilayer distance $\bar{d}$ is an additional variational parameter.
Among the five afore-mentioned structures, the one which provides 
the minimum of the energy $E$ is chosen as stable.

{The appearance of three, four or more layered structures cannot
be, in principle, excluded.
Other approaches, like Monte Carlo (MC) simulations, are more suitable 
for detecting such complicated structures.
We only know that the case $\Delta=0$ and the two extreme limits $d\to 0$ 
(structure I) and $d\to\infty$ (structure V) for all $\Delta\in [0,1]$ 
are treated correctly within our assumption:

\begin{itemize}
\item
As soon as $\Delta>0$, the stability along $z$-axis of a particle inside 
structure I is influenced ``positively'' by image/self-image effects 
and ``negatively'' by direct Coulomb interaction with all other particles.  
The image effects become stronger when the distance between walls $d$
decreases; there always exists a finite distance $d$ when the image force
overwhelms the direct repulsion by other particles.
This is why structure I is expected to be stable in a finite interval
of $d$ values, in contrast to the $\Delta=0$ case.

\item
In the opposite limit $d\to\infty$, our energy per particle (\ref{energy}) 
reduces to that of the hexagonal Wigner crystal with images
for one-wall geometry. 
Indeed, in that limit, the following terms survive from the regularized
version of (\ref{latticesums}):
the sum $\sum_{j\ne 1} 1/R_{1j}^{AA}$, $\Delta/[\varepsilon(d-\bar{d})]$ from 
the $n=1$ term of $u_{\rm im}(0;\bar{d}/2,\bar{d}/2)$ and
$$
\frac{\Delta}{\varepsilon} \left[ \sum_{j\ne 1} 
\frac{1}{\sqrt{(R_{1j}^{AA})^2+(d-\bar{d})^2}} + 2\pi\sigma (d-\bar{d})
\right]
$$ 
from the $n=1$ term of $\sum_{j\ne 1} u_{\rm im}(R_{1j}^{AA};\bar{d}/2,\bar{d}/2)$.
Substituting these terms into Eq. (\ref{energy}) and expressing everything
in terms of the distance $l$ of the hexagonal Wigner crystal from the wall, 
we recover the basic expression for the energy per particle (9) in 
\cite{Samaj12c} for the charge valence $q=1$.
The two Wigner crystals form, together with the corresponding charged wall,
neutral entities shifted as in structure V to ensure the lowest energy.
\end{itemize}}

We shall work with dimensionless distances
\begin{equation}
\eta = d \sqrt{\sigma} , \qquad \bar{\eta} = \bar{d} \sqrt{\sigma} , \qquad
\xi = l \sqrt{\sigma} ,
\end{equation}
such that $\xi = (\eta-\bar{\eta})/2$.
In the asymptotic limit $\eta\to\infty$, the dimensionless distance of 
the Wigner layer from the corresponding wall is given by   
\begin{equation} \label{xi}
\xi^* = \frac{t}{3^{1/4}\sqrt{2}}  \qquad (\eta\to\infty) ,
\end{equation}
where $t$ is the $\Delta$-dependent solution of Eqs. (\ref{t}) and (\ref{tt}).
This formula provides a check of the accuracy of our results in the region of 
structure V as well as criterion for reaching the asymptotic one-wall regime.

\section{Numerical results}

\begin{figure}[!ht]
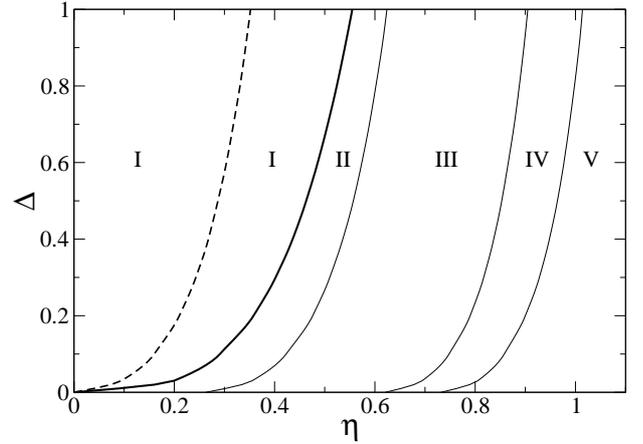

\begin{center}
\onefigure[width=0.45\textwidth]{Fig3.eps}
\caption{{The phase diagram in the $(\eta,\Delta)$ plane.
The dashed line, lying entirely in the structure-I region, corresponds 
to the absolute ground state (minimum of the total energy).}}
\label{fig:phasediagram}
\end{center}
\end{figure}

\begin{table}[hbt]
\caption{The values of the dimensionless layer separation $\bar{\eta}$
at phase transitions between structures ${\rm I\to II\to III\to IV\to V}$, 
for a given dielectric jump $\Delta$.}
\label{Table1}
\begin{center}
\begin{tabular}{lccccc}
$\Delta$ & $\bar{\eta}_{\, \rm I\to II}$ & $\bar{\eta}_{\, \rm II\to III}$ 
& $\bar{\eta}_{\, \rm III\to IV}$ & $\bar{\eta}_{\, \rm IV\to V}$ \\ & & & & & \\
0.2 & 0 & 0.273 & 0.619 & 0.732  \\
0.4 & 0 & 0.267 & 0.607 & 0.723  \\
0.6 & 0 & 0.259 & 0.596 & 0.712  \\
0.8 & 0 & 0.252 & 0.587 & 0.703  \\
1.0 & 0 & 0.245 & 0.578 & 0.693 
\end{tabular}
\end{center}
\end{table}

We have performed numerical calculations using the Mathematica software.
{The phase diagram in the $(\eta,\Delta)$ plane is pictured in
fig. \ref{fig:phasediagram}.}
Generally speaking, the transition values of $\eta$ exhibit a much stronger
dependence on $\Delta$ than the ones of $\bar{\eta}$ presented in 
table \ref{Table1}.
The relevant fact is that for every $\Delta>0$ the particles form the Wigner 
monolayer ($\bar{\eta}=0$) with structure I up to some positive 
$\eta_{\,\rm I\to II}>0$, as was anticipated.

\begin{figure}[!ht]
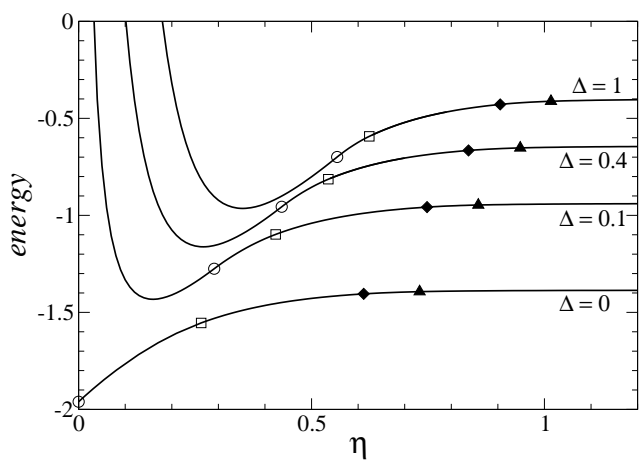

\begin{center}
\onefigure[width=0.46\textwidth]{Fig4.eps}
\caption{The dimensionless ground-state energy per particle, 
$E\varepsilon/(Ne^2\sqrt{2\sigma})$, versus the dimensionless distance 
between the walls $\eta$, for four values of the dielectric jump 
$\Delta=0$, $0.1$, $0.4$ and $1$. The
symbols represent the transition points between structures: 
open circle for ${\rm I\to II}$, open square for ${\rm II\to III}$, 
filled diamond for ${\rm III\to IV}$ and filled triangle for ${\rm IV\to V}$.}
\label{fig:energy}
\end{center}
\end{figure}

Interestingly, for $\Delta>0$ the ground-state energy is not 
a monotonic function of $\eta$. 
It reaches its global minimum at some $\eta_{\min}$ which is smaller than 
$\eta_{\, \rm I\to II}$, i.e. the particles form the structure-I Wigner monolayer
at this point, see fig. \ref{fig:energy}.
Such a behavior differs substantially from the homogeneous case $\Delta=0$, 
for which the energy grows monotonically. 

\begin{figure}[!ht]
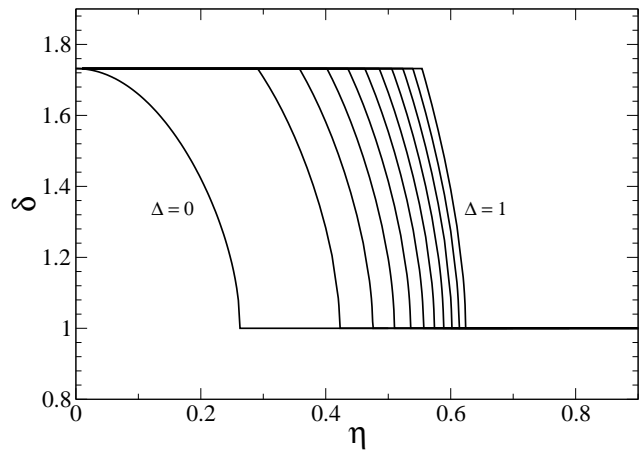

\begin{center}
\onefigure[width=0.46\textwidth]{Fig5.eps}
\caption{The aspect ratio $\delta$ of structures I-III versus 
the dimensionless distance between the walls $\eta$.
The dielectric jump $\Delta$ ranges between 0 to 1, from left to right 
with step $0.1$.
$\delta=\sqrt{3}=1.73205\ldots$ for structure I, $1<\delta<\sqrt{3}$ within 
structure II and $\delta=1$ for structure III.}
\label{fig:aspectratio}
\end{center}
\end{figure}

At point $\eta_{\,\rm I\to II}$, the particle system starts to bifurcate to 
a Wigner bilayer ($\bar{\eta}>0$) and at the same time structure I
evolves into structure II with the varying aspect ratio 
$1<\delta<\sqrt{3}$ (see fig. \ref{fig:aspectratio}).  
Structure II ends up at the second-order transition point 
$\eta_{\,\rm II\to III}$, where structure III with the fixed aspect ratio 
$\delta=1$ determines the ground state energy.

\begin{figure}[!ht]
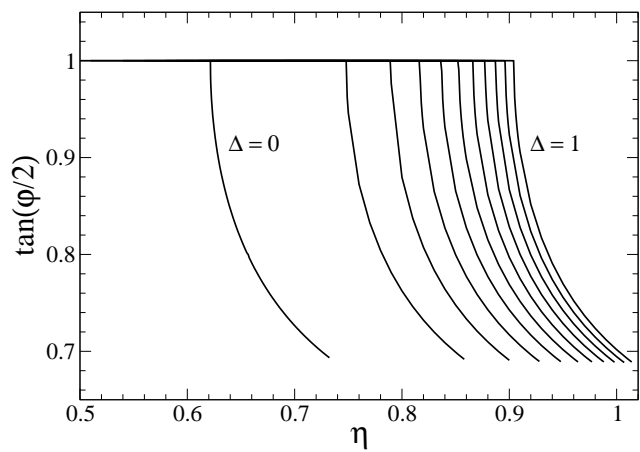

\begin{center}
\onefigure[width=0.46\textwidth]{Fig6.eps}
\caption{Dependence of the angle $\varphi$ for structure IV on 
the dimensionless distance between the walls $\eta$.
The dielectric jump $\Delta$ ranges between 0 to 1, from left to right 
with step $0.1$.
Each line ends at the transition point $\eta_{\,\rm IV\to V}$.}
\label{fig:angle}
\end{center}
\end{figure}

The second-order phase transition from structure III $(\varphi=\pi/2)$
to structure IV, with a varying angle $\varphi$ such that $\tan(\varphi/2)<1$, 
takes place at point $\eta_{\,\rm III\to IV}$.
The plots of the order parameter $\tan(\varphi/2)$ versus the dimensionless
inter wall distance are pictured for the dielectric jumps 
$\Delta=0, 0.1,\ldots, 1$ in fig. \ref{fig:angle}. 

\begin{figure}[!ht]
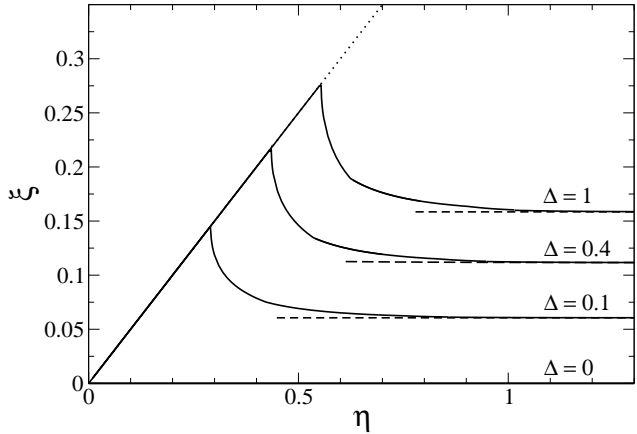

\begin{center}
\onefigure[width=0.46\textwidth]{Fig7.eps}
\caption{Dimensionless distance of the Wigner sub-layer from 
the nearest wall, $\xi$, versus the dimensionless distance between 
the walls $\eta$, for four values of the dielectric jump 
$\Delta=0$, $0.1$, $0.4$ and $1$. The dotted line has slope $1/2$.
For $\Delta=0$, we have the trivial result $\xi=0$, i.e. particles
collapse onto the wall surface, as required by Earnshaw theorem. 
The horizontal dashed lines correspond to the asymptotic $\eta\to\infty$ 
values for the one-wall problem $\xi^*$, given by (\ref{xi}).} 
\label{fig:distance}
\end{center}
\end{figure}

At the first-order transition point $\eta_{\,\rm IV\to V}$, the angle $\varphi$ 
skips from the value $\sim 1.21$ within structure IV, which is practically 
independent of $\Delta$, to $\pi/3\sim 1.05$ for structure V. 
An interesting question is whether also the layer separation $\bar{\eta}$ 
exhibits a skip at $\eta_{\,\rm IV\to V}$. 
The values of $\bar{\eta}_{\rm IV\to V}$ for various $\Delta$'s in the last 
column of table \ref{Table1} turn out to be the same for 
both structures IV and V at the transition point, 
within our number specification to 3 decimal digits. 

In the last fig. \ref{fig:distance}, we plot the dimensionless
distance of the Wigner sub-layer from the nearest wall, $\xi$, versus 
$\eta$, for four values 
of the dielectric jump $\Delta\in [0,1]$.
The dielectric homogeneous case $\Delta=0$ with the trivial $\xi=0$,
i.e. particle collapse onto the wall surface, is shown for comparison.
For the corresponding transition points between the structures,
see fig. \ref{fig:energy}.
If $\Delta>0$, the Wigner monolayer with structure I stays at $z=0$ up to 
$\eta_{\,\rm I\to II}$ and so $\xi=\eta/2$ for $0\le \eta < \eta_{\,\rm I\to II}$.
Interestingly, the value of $\xi$ attained at $\eta_{\,\rm I\to II}$ 
is always larger than the asymptotic $\eta\to\infty$ value $\xi^*$ 
for the one-wall problem, given by Eq. (\ref{xi}) and depicted
by horizontal dashed lines in fig. \ref{fig:distance}. 
This is why, after bifurcation of the Wigner monolayer to the series
of bilayer structures II-V by increasing $\eta$, 
the distance $\xi$ goes to $\xi^*$ from above. 
For $\eta\sim 1.3$, $\xi$ coincides with the asymptotic value $\xi^*$ 
to within 4 decimal digits.

\section{Conclusion}
We have extended the standard study of the homogeneous dielectric version of 
the ground state problem for particles with Coulomb pair 
interactions, constrained between two equivalently charged wall, 
to a more realistic inhomogeneous dielectric situation with 
repulsive electrostatic image charges.
The analysis of the additional Coulomb lattice sums due to the images 
of particles and surface charges was performed. 
All lattice sums are treated within the recent method 
\cite{Samaj12a,Samaj12b} which uses a bilayer generalization of
Misra functions (\ref{znu}) to construct very quickly convergent
series expansions for the ground state energy.
The  numerical results obtained show that inclusion of repulsive images 
has two fundamental effects.
Firstly, the particles do not collapse onto the charged surfaces of
the walls, but create a Wigner monolayer or bilayer symmetrically
with a nonzero spacing from the walls.
The Wigner monolayer with the hexagonal structure I remains stable up 
to a certain (dimensionless) distance between the walls $\eta_{\,\rm I\to II}$, 
where it starts to bifurcate into the series of bilayer structures II-V.
Secondly, the ground state energy exhibits as the function of the
distance between the walls a minimum at some $\eta_{\min}$ lying in
the region of the structure-I Wigner monolayer. 

{Although thermal fluctuations are supposed to destroy a classical 
Wigner crystal at very low temperatures, their fingerprints persist 
within a recent
strong-coupling theory \cite{Samaj11} to higher (room) temperatures,
in agreement with MC simulations.
Based on the present work, we plan to extend that strong-coupling
study by including wall images.} 

\acknowledgments
L. \v{S}. is grateful to LPTMS for its hospitality. 
The support received from Grants VEGA No. 2/0049/12 and CE-SAS QUTE
is acknowledged.

\end{document}